\documentclass[12pt]{article}

\begin{document}
\title{Exact uncertainty approach in quantum mechanics and quantum gravity}
\author{Michael J. W. Hall\\
Theoretical Physics, IAS\\ Australian National
University\\ Canberra ACT 0200, Australia}
\date{}
\maketitle

%If the article contains enunciations (e.g., theorems, lemmas, etc., 
%please use the following
%macros which are available in LaTeX:
%\newtheorem{theorem}{Theorem}[section]
%\newtheorem{lemma}[theorem]{Lemma}
%etc....

\begin{abstract}
The assumption that an ensemble of classical particles is subject to nonclassical momentum fluctuations, with the fluctuation uncertainty fully determined by the position uncertainty, has been shown to lead  from the classical equations of motion to the 
Schr\"{o}dinger equation.  This `exact uncertainty' approach may be generalised to ensembles of gravitational fields, where nonclassical fluctuations are added to the field momentum densities, of a magnitude determined by the uncertainty in the metric tensor components.  In this way one obtains the Wheeler-DeWitt equation of quantum gravity, with the added bonus of a uniquely specified operator ordering.  No {\it a priori} assumptions are required concerning the existence of wavefunctions, Hilbert spaces, Planck's constant, linear operators, etc.  Thus this approach has greater transparency than the usual canonical approach, particularly in regard to the connections between quantum and classical ensembles.  Conceptual foundations and advantages are emphasised.

\end{abstract}

%USE FOR REVTEX
%\pacs{03.65.Ta}
%\maketitle

\newpage

\section{Introduction}

Uncertainty is a fundamental element of quantum mechanics: states cannot simultaneously assign small uncertainties to every observable; measuring one observable changes the uncertainties of other observables; vacuum uncertainties lead to macroscopic phenomena such as the Casimir effect.

The aim of the exact uncertainty approach is to show that uncertainty may be promoted to {\it the} fundamental element distinguishing classical and quantum mechanics.  In this approach nonclassical fluctuations are added to the usual deterministic connection between the configuration and momentum properties of a physical system.  Making the postulate that the uncertainty introduced to the momentum (i.e., the fluctuation strength) is fully determined by the uncertainty in the configuration (i.e., by the configuration probabilty density) then leads from the classical to the quantum equations of motion 
\cite{hallreg, hkr}.  

The exact uncertainty approach has certain advantages over the canonical approach (eg, there are no assumptions or postulates required regarding the existence of wavefunctions, Hilbert spaces, and operator orderings), and even its apparent limitations are of interest (the fluctuation strength must contribute quadratically to the energy for the approach to go through, which is in fact guaranteed for all known particles and bosonic fields).  The approach is thus rather economical, in that it appears to provide the bare minimum that is necessary for the description of physical quantum systems.

This review begins with a careful discussion of the exact uncertainty approach for particles in section 2.  This focuses attention on the main ideas underlying the approach, and in section 3 allows for the separation of these ideas from some of the distracting technical complications that arise in their application to gravitational fields.  Benefits of the exact uncertainty approach are discussed in section 4, as well as some interpretational issues.

\section{Exact uncertainty approach: particles}

\subsection{Black magic and the Schr\"{o}dinger equation}

In the canonical quantisation procedure for a spinless particle, one begins with the classical Hamiltonian \begin{equation} \label{ham} 
H(x,p,t)= \frac{p^2}{2m}+V(x) . 
\end{equation}
One then introduces a `wavefunction' $\psi(x,t)$, residing in the Hilbert space of square-integrable functions on configuration space, and a constant, $\hbar$, having units of action, and writes down the linear equation of motion
\begin{equation} \label{se}
i\hbar \frac{\partial\psi}{\partial t} = H(x, -i\hbar \nabla, t) \psi =
\frac{-\hbar^2}{2m} \nabla^2 \psi + V\psi.
\end{equation}
The wavefunction is given physical import via the statistical prediction that the probability density for a position measurement is given by 
\begin{equation}  \label{prob}
P(x,t)=|\psi(x,t)|^2 . 
\end{equation}

The above recipe immediately raises a number of questions, especially for the student meeting quantum mechanics for the first time.  Where do $\psi$, $\hbar$ and $i$ come from ?  Why write down a linear operator equation ?  How does probability suddenly appear ?  The more advanced student may also worry about operator ordering problems - for example, if the mass $m$ is position dependent, how does one decide between $m^{-1}p^2$, $pm^{-1}p$, etc, to get the `right' Schr\"{o}dinger equation ? 

Thus, in the canonical approach, there appear to be many elements of `black magic' involved in moving from a classical to a quantum description.  Of course it can be argued that this is not a fundamental difficulty, as quantum mechanics must in the end be justified by its physical applicability, rather than by any particular relationship with classical mechanics.  Even so, this relationship is so tenuous in the canonical formulation that it is little wonder that quantum mechanics is so difficult to understand, even {\it before} one gets to the measurement problem, Bell inequalities, etc.

In the following subsections a rather different approach is taken to obtaining the Schr\"{o}dinger equation (\ref{se}), in which the connection between the quantum and classical equations of motion is much stronger, and where the primary difference may be understood in terms of statistical uncertainty.

\subsection{Ensembles on configuration space}

The core concept in the exact uncertainty approach to quantisation is statistical uncertainty.  Rather than bringing in probability as an interpretational follow-up to the {\it ad hoc} introduction of a wavefunction, probability is instead built into the approach from the very beginning, and wavefunctions and the Schr\"{o}dinger equation are derived rather than postulated.

In particular, consider the possibility that, whether for theoretical or practical purposes, the {\it position} of a particle is an inherently imprecise physical notion.  Thus, to describe position, one must introduce a probability density, $P(x,t)$, on configuration space, and deal with a corresponding {\it statistical ensemble} of particles.

To describe the dynamics of such ensembles, it will be assumed that the evolution of the fundamental quantity, $P(x,t)$, follows from an action principle.  This immediately requires the existence of some conjugate function $S(x,t)$ on configuration space, and a corresponding {\it ensemble Hamiltonian} $\tilde{H}[P,S,t]$, such that the equations of motion are generated by the action principle $\delta A = 0$, with 
\cite{goldstein}
\begin{equation} \label{action}
   A = \int dt\, \left[ \tilde{H} + \int dx\, P \frac{\partial S}{ \partial t} \right] .
\end{equation}
Note that one could, of course, deal with a Lagrangian rather than a Hamiltonian formalism, but the latter has some technical advantages 
\cite{hallreg} and so will be used here.  

The corresponding equations of motion for $P$ and its conjugate $S$ have the Hamiltonian form
\begin{equation} \label{conj}
  \frac{\partial P}{\partial t} = \frac{\delta \tilde{H}}{\delta S},~~~~~~~~ \frac{\partial S}{\partial t} = -\frac{\delta \tilde{H}}{ \delta P} 
\end{equation}
as a consequence of Eq.~(\ref{action}), where $\delta/\delta f$ denotes the functional derivative with respect to function $f$ \cite{goldstein}.  It may be shown that conservation of probability requires that the ensemble Hamiltonian $\tilde{H}$ is invariant under $S\rightarrow S+{\rm constant}$ \cite{super}.  Note also that if $\tilde{H}$ has no explicit time dependence then its value is a conserved quantity \cite{goldstein}, corresponding to the energy of the ensemble.

As a particular example of interest, consider the `classical' ensemble Hamiltonian defined by 
\begin{equation} \label{hamc}
\tilde{H}_c[P,S] := \int dx\, P\left[ \frac{|\nabla S|^2}{2m} + V \right]
\end{equation}
for some constant $m$ and function $V(x)$.  Note that this is the simplest possible form for $\tilde{H}$ that is scalar, linear in $P$, and which conserves probability. The corresponding equations of motion follow via Eqs.~(\ref{conj}) as
\begin{equation} \label{conthj}
\frac{\partial P}{\partial t}+\nabla .\left[P\frac{\nabla S}{m} \right] =0,~~~~~~ \frac{\partial S}{\partial t} + \frac{|\nabla S|^2}{2m}+V=0 .
\end{equation}
The first of these may be recognised as a continuity equation for an ensemble of particles having flow velocity $v=m^{-1}\nabla S$, and the second as the Hamilton-Jacobi equation for a classical particle of mass $m$ moving in a potential $V$ \cite{goldstein}. Note also that 
$\tilde{H}_c$ indeed corresponds to the average energy of a classical ensemble, providing that $\nabla S$ is interpreted as the momentum of a particle at position $x$.

The above formalism, based on an action principle for the position probability density, thus successfully describes the motion of ensembles of classical particles.  However, it is considerably more general in its applicability (see \cite{super} for a review).  In particular, the essential difference between {\it classical} and {\it quantum} ensembles becomes a `matter of form', being characterised by a simple difference in the forms of the corresponding ensemble Hamiltonians, $\tilde{H}_c$ and $\tilde{H}_q$.  In the `exact uncertainty' approach to be followed here, this makes the technical leap from classical to quantum motion relatively small in comparison with the canonical approach.  It will be argued that the conceptual leap is also smaller, being based on a single nonclassical concept: the addition of nonclassical momentum fluctuations.

\subsection{Adding momentum fluctuations}

As noted above, $\tilde{H}_c$ in Eq.~(\ref{hamc}) corresponds to the usual energy of a classical ensemble if it is assumed that $\nabla S$ is the particle momentum associated with position $x$.  However, given that position has been taken to be an inherently imprecise notion, requiring a statistical ensemble for its description, the classical assumption of a deterministic connection between position and momentum is not necessarily warranted or justifiable.  Hence this connection will be relaxed to the assumption that the physical momentum is of the form
\begin{equation} \label{fluc}
p = \nabla S + f ,
\end{equation}
where the fluctuation field $f$ vanishes everywhere on average.  Note that such fluctuations introduce indeterminism at the level of individual particles, as they remove the possibility of integrating $v=m^{-1}\nabla S$ to obtain particle trajectories.  However, as will be seen, these fluctuations need not remove determinism at the ensemble level.

An overline will be used to denote averaging over the fluctuations at a given position. Thus $\overline{f}=0$ and $\overline{p}=\nabla S$ by assumption, and the classical ensemble energy is replaced by
\begin{eqnarray}
< E> &=& \int dx\, P \left[ (2m)^{-1}\overline{|\nabla S + f|^2} + V \right] \nonumber\\
&=& \int dx\, P \left[ (2m)^{-1}(|\nabla S|^2+2 \overline{f}.\nabla S + 
\overline{f.f}) + V\right] \nonumber\\ \label{efluc}
&=& \tilde{H}_c + \int dx\, P\, \frac{\overline{f.f}}{2m} .
\end{eqnarray}
Thus the classical energy is increased by the average kinetic energy of the momentum fluctuations.

The question now is whether this modified classical ensemble, incorporating nonclassical momentum fluctuations as per Eq.~(\ref{fluc}), can be subsumed within the general formalism of the previous section.  The answer is clearly yes, provided that the average strength of the fluctuations at each point, $\overline{f.f}$ in Eq.~(\ref{efluc}), is determined by some function of $P$, $S$ and their derivatives, i.e.,
\begin{equation} \label{alpha} 
\overline{f.f} = \alpha(x,P, S, \nabla P, \nabla S,\dots) , 
\end{equation}
where the dots denote possible higher derivatives of $P$ and $S$.
For one can then define a modified ensemble Hamiltonian
\begin{equation} \label{hammod}
\tilde{H}_q[P,S] := \tilde{H}_c[P,S] + \int dx\, P\, 
\frac{\alpha(x,P,S, \nabla P, \nabla S,\dots )}{2m}
\end{equation}
consistent with Eqs.~(\ref{fluc}) and (\ref{efluc}), and which reduces to the classical ensemble Hamiltonian in the limit of vanishing fluctuations.

The aim of the exact uncertainty approach is to fix the form of the function $\alpha$, and hence that of $\tilde{H}_q$, uniquely.  This is done by requiring three generally desirable principles to be satisfied (causality, invariance and independence), as well as an `exact uncertainty' principle.  The resulting equations of motion for $P$ and $S$ are equivalent to the Schr\"{o}dinger equation for a quantum ensemble of particles.

\subsection{Obtaining the Schr\"{o}dinger equation}

Four suitable principles for determining the function $\alpha$ in 
Eqs.~(\ref{alpha}) and (\ref{hammod}) have been discussed in 
Refs.~\cite{hallreg} and \cite{fort}.  Slightly different versions are given here, based on those used for bosonic fields in Ref.~\cite{hkr}, so as to make a more direct connection with quantum gravity.  

The first three principles are very natural on physical grounds, and hence are not considered particularly restrictive.  They require that (i) {\it the modified ensemble Hamiltonian $\tilde{H}_q$ in 
Eq.~(\ref{hammod}) leads to causal equations of motion} (thus $\alpha$ can cannot depend on second and higher derivatives of $P$ and $S$);  (ii) {\it the respective fluctuation strengths for non-interacting uncorrelated ensembles are independent} (thus 
$\overline{f_1.f_1}$ and $\overline{f_2 .f_2}$ are independent of $P_1$ and $P_2$ respectively, when $P(x)=P_1(x_1)P_2(x_2)$); and (iii) {\it the fluctuations transform correctly under linear canonical transformations} (thus, noting Eq.~(\ref{fluc}), $f\rightarrow L^Tf$ for any invertible linear coordinate transformation $x\rightarrow L^{-1}x$). 

The fourth and final principle is, in contrast, rather more restrictive, requiring that (iv) {\it the strength of the momentum fluctuations, $\alpha=\overline{f.f}$, is determined solely by the uncertainty in position} (thus, since $P$ contains all available information about the position uncertainty, $\alpha$ can in fact only depend on $x$, $P$ and its derivatives).  This is referred to as the {\it exact uncertainty principle} \cite{hallreg,hkr,fort}, since it postulates a precise (but unspecified!) relation between the fluctuation statistics and the position statistics.  
 
As has been shown elsewhere \cite{hallreg,hkr, fort}, the above four principles lead to the unique form
\begin{equation} \label{hamq}
\tilde{H}_q[P,S] = \tilde{H}_c[P,S] + C \int dx\, 
\frac{\nabla P.\nabla P}{2mP} ,
\end{equation}
where $C$ is a positive universal constant (i.e., having the same value for all ensembles).  Moreover, if one now {\it defines} 
\[ \hbar:= 2\sqrt{C},~~~~~~\psi:= \sqrt{P}e^{iS/\hbar}, \]
it is straightforward to show that the equations of motion for $P$ and $S$ are equivalent to the Schr\"{o}dinger equation (and its conjugate) in Eq.~(\ref{se}). 

Note that the four principles used above may all be expressed in terms of the statistics of the nonclassical momentum fluctuations.  Thus the amount of `black magic' required to obtain the Schr\"{o}dinger equation is reduced to a {\it single} nonclassical element, in contrast to the many ingredients needed in the canonical approach.  Further comparisons are made in section 4.

\section{Exact uncertainty approach: gravity}

\subsection{Ensembles of gravitational fields}

The general framework of the exact uncertainty approach has been carefully laid out in the previous section.  Its application to quantum gravity is conceptually very much the same as for quantum particles, but account must be taken of technical complications such as replacing probability functions by probability functionals, and dealing with constraints. Technical details have been discussed in Ref.~\cite{hkr} (which deals with bosonic fields in general), and hence it is the conceptual elements which will be emphasised here.

The configuration of a classical gravitational field is determined by the corresponding metric
\[ ds^2 = -(N^2-h^{ij}N_iN_j)dt^2 +2N_idx^idt+h_{ij}dx^idx^j \]
in spacetime, where $N$ and $N_i$ are the lapse and shift functions, and $h_{ij}$ is the spatial 3-metric.  Since the lapse and shift functions may be specified arbitrarily (reflecting invariance under arbitrary coordinate transformations), the configuration space of the field is the set of spatial 3-metrics $\{h_{ij}\}$.

As for the case of particles in section 2, consider now the possibility that the configuration of the field is an inherently imprecise notion, hence requiring a probability functional, $P[h_{ij}]$, for its description.  It will again be assumed that the dynamics of the corresponding statistical ensemble are generated by an action principle, $\delta A=0$, with
\[ A = \int dt\, \left[ \tilde{H} + \int Dh\, P \frac{\partial S}{ \partial t} \right]  \]
analogous to Eq.~(\ref{action}).  Here $\int Dh$ denotes the functional integral over the configuration space, and the ensemble Hamiltonian $\tilde{H}$ is dependent on $P[h_{ij}]$ and its conjugate functional 
$S[h_{ij}]$.  It follows that the equations of motion for the ensemble have the Hamiltonian form
\begin{equation} \label{canon}
  \frac{\partial P}{\partial t} = \frac{\Delta \tilde{H}}{\Delta S},~~~~~~~~ \frac{\partial S}{\partial t} = -\frac{\Delta \tilde{H}}{ \Delta P}  ,
\end{equation}
where $\Delta/\Delta F$ denotes the variational derivative with respect to functional $F$ \cite{hkr}.

A suitable `classical' ensemble Hamiltonian may be constructed from knowledge of the classical equations of motion for an individual field, and is given by the functional integral \cite{hkr}
\begin{equation} \label{hamgc}
\tilde{H}_c[P,S] = \int Dh\, P H_0[h_{ij}, \delta S/\delta h_{ij}],
\end{equation}
where the functional $H_0$ has the form
\begin{equation} \label{h0}
H_0[h_{ij},\pi^{ij}] := \int dx\, \left[ N\left( \frac{1}{2} G_{ijkl} 
\pi^{ij}\pi^{kl} + V(h_{ij})\right) - 2N_i\pi^{ij}_{~|j} \right] .
\end{equation}
Here $G_{ijkl}$ is the Wheeler-DeWitt supermetric, $V$ is the negative of twice the product of the 3-curvature scalar with $(\det h)^{1/2}$, and $|j$ denotes the covariant 3-derivative.  Eq.~(\ref{h0}) may be recognised as the {\it single-field} Hamiltonian \cite{dewitt}.

It may be checked that the equations of motion corresponding to the ensemble Hamiltonian $\tilde{H}_c$ in Eq.~(\ref{hamgc}) are given by
\[ \frac{\partial P}{\partial t}+\int dx\, \frac{\delta}{\delta h_{ij}} \left( P\dot{h}_{ij}\right)=0,~~~~~ \frac{\partial S}{\partial t} + H_0[h_{ij},\delta S/\delta h_{ij}] =0,  \]
where one defines 
\[ \dot{h}_{ij}:= NG_{ijkl}\frac{\delta S}{\delta h_{kl}} - N_{i|j}- 
N_{j|i} .  \]
These equations of motion correspond to the conservation of probability with probability flow $\dot{h}_{ij}$, and the Hamilton-Jacobi equation for an individual gravitational field with configuration $h_{ij}$ 
\cite{peres, gerlach, rovelli}.

As is well known, the lack of conjugate momenta for the lapse and shift components $N$ and $N_i$ of the metric places constraints on the classical equations of motion \cite{dewitt}.  In the ensemble formalism these constraints take the form \cite{hkr}
\begin{eqnarray} \label{conp}
\frac{\delta P}{\delta N}=\frac{\delta P}{\delta N_i}=\frac{\partial P}{\partial t}=0,&~&~~~~~\left(\frac{\delta P}{\delta h_{ij}}\right)_{|j}=0,\\ \label{cons}
\frac{\delta S}{\delta N}=\frac{\delta S}{\delta N_i}=\frac{\partial S}{\partial t}=0,&~&~~~~~\left(\frac{\delta S}{\delta h_{ij}}\right)_{|j}=0 ,
\end{eqnarray}
and correspond to invariance of the dynamics with respect to the choice of lapse and shift functions and initial time, and to the invariance of $P$ and $S$ under arbitrary spatial coordinate transformations.

Applying these constraints to the above classical equations of motion yields, for the `Gaussian' choice $N=1$, $N_i=0$ of lapse and shift functions, the reduced classical equations
\begin{equation} \label{reduced} 
\frac{\delta}{\delta h_{ij}} \left( PG_{ijkl} \frac{\delta S}{ 
\delta h_{kl}} \right)=0,~~~~~~\frac{1}{2}G_{ijkl} \frac{\delta S}{ \delta h_{ij}} \frac{\delta S}{ \delta h_{kl}}  + V = 0 
\end{equation}
for $P$ and $S$.

\subsection{Obtaining the Wheeler-DeWitt equation}

The exact uncertainty approach of sections 2.3 and 2.4 may now be followed in a straightforward manner to obtain a modified ensemble Hamiltonian that generates the quantum equations of motion.  It is first assumed that the classical deterministic relation between the field configuration $h_{ij}$ and its conjugate momentum density 
$\pi^{ij}$ is relaxed to
\[ \pi^{ij} = \frac{\delta S}{\delta h_{ij}} + f^{ij} , \]
analogous to Eq.~(\ref{fluc}), where $f^{ij}$ vanishes on average for all configurations.  This adds a kinetic term to the average ensemble energy analogous to Eq.~(\ref{efluc}), with
\[ \tilde{H}_q:= \langle E\rangle = \tilde{H}_c + \frac{1}{2}\int Dh\, P\int dx\, N G_{ijkl} \overline{f^{ij}f^{kl}} \]
(note that the term in Eq.~(\ref{h0}) linear in the derivative of 
$\pi^{ij}$ can be integrated by parts to give a term directly proportional to $\pi^{ij}$, which remains unchanged when the fluctuations are added and averaged over).

One now fixes the form of the the modified Hamiltonian $\tilde{H}_q$ using precisely the same principles of causality, independence, invariance and exact uncertainty used in section 2.4 (see 
Ref.~\cite{hkr} for technical details), leading to the result
\begin{equation} \label{hamqg}
\tilde{H}_q[P,S] = \tilde{H}_c[P,S] + \frac{C}{2}\int Dh \int dx\, N
G_{ijkl} \frac{1}{P}\frac{\delta P}{\delta h_{ij}} 
\frac{\delta P}{\delta h_{kl}},
\end{equation}
analogous to Eq.~(\ref{hamq}), where $C$ is a positive universal constant (i.e., with the same value for all fields).  

The corresponding modified equations of motion may be calculated via 
Eqs.~(\ref{canon}), and the constraints in Eqs.~(\ref{conp}) and 
(\ref{cons}) applied to obtain reduced equations analogous to 
Eq.~(\ref{reduced}).  If one now {\it defines}
\[  \hbar:=2\sqrt{C},~~~~~~~~~\Psi[h_{ij}]:=\sqrt{P}e^{iS/\hbar},\]
these reduced equations can be rewritten in the form \cite{hkr}
\begin{equation} \label{wdw}
\left[ -\frac{\hbar^2}{2} \frac{\delta}{\delta h_{ij}} G_{ijkl} 
\frac{\delta}{\delta h_{kl}} + V \right] \Psi = 0 .
\end{equation}
which may be recognised as the Wheeler-DeWitt equation for quantum gravity.  Note further that the constraints in Eqs.~(\ref{conp}) and (\ref{cons}) may be rewritten in terms of the wavefunctional $\Psi$ as
\begin{equation} \label{conwdw}
\frac{\delta \Psi}{\delta N}=\frac{\delta \Psi}{\delta N_i}=
\frac{\partial \Psi}{\partial t}=0,~~~~~\left(\frac{\delta \Psi}{ \delta h_{ij}}\right)_{|j}=0 .
\end{equation}

One very interesting aspect of the Wheeler-DeWitt equation derived in 
Eq.~(\ref{wdw}) is that it has been obtained with a particular operator ordering: the supermetric $G_{ijkl}$ is sandwiched between the two functional derivatives.  This contrasts with the canonical approach, which is unable to specify a unique ordering \cite{dewitt}.  It should be noted that different orderings can lead to different physical predictions \cite{wiltshire}, and hence the exact uncertainty approach is able to remove ambiguity in this respect.  An analogous removal of ambiguity is obtained for quantum particles having a
position-dependent mass \cite{hkr}, with the exact uncertainty approach specifying, via Eq.~(\ref{hamq}), the unique `sandwich' ordering
\begin{equation} \label{order}
i\hbar \frac{\partial\psi}{\partial t} =
\frac{-\hbar^2}{2} \nabla . \frac{1}{m}\nabla \psi + V\psi
\end{equation}
for the corresponding Schr\"{o}dinger equation.

\section{Discussion}

It has been shown that uncertainty can be taken as the conceptual basis for quantisation, for both particles and gravitational fields.  Physical ensembles are described by a probability density on configuration space, $P$; a corresponding conjugate quantity $S$; and an ensemble Hamiltonian $\tilde{H}[P,S]$.  The transition from classical ensembles to quantum ensembles then follows as a consequence of the addition of nonclassical momentum fluctuations, under the assumption that the fluctuation uncertainty is fully determined by the configuration uncertainty.  

In contrast to the canonical approach, the Schr\"{o}dinger and Wheeler-DeWitt equations are {\it derived} in the exact uncertainty approach, rather than simply postulated.  Further, while the canonical approach requires the {\it ad hoc} postulate of a statistical connection between probability and the wavefunction, as per Eq.~(\ref{prob}),  in the exact uncertainty approach this connection is a simple consequence of the definition of $\psi$ in terms of $P$ and $S$.  

Note that no assumptions concerning Hilbert spaces, linearity, superposition, entanglement, wavefunctions, operator-ordering or the like are used or required to obtain the quantum ensemble Hamiltonian (and hence the quantum equations of motion).  Moreover, Planck's constant appears as a consequence of a derived universal scale for the nonclassical momentum fluctuations, as opposed to being an unexplained constant multiplying time and configuration derivatives in the canonical approach.  It is concluded that the exact uncertainty approach provides, both conceptually and technically, a far more constructive approach to quantum mechanics.

Another point of interest, in regard to the comparison between the canonical and exact uncertainty approaches, is the apparent limitation of the latter to a restricted class of systems.  In particular, the momentum of a classical ensemble must contribute {\it quadratically} to the ensemble energy for the exact uncertainty approach to go through, whereas the canonical approach is indifferent in this regard.  Fortunately, however, all nonrelativistic particles and all relativistic integer-spin fields fall within this `quadratic' class (essentially as a consequence of the fundamental equation of motion for the configuration being second order in time, as per Newton's second law), and hence all such systems are covered by the exact uncertainty approach as described above (fermionic fields will be discussed elsewhere).  Another example of physical economy is the removal of operator-ordering ambiguities, such as in Eqs.~(\ref{wdw}) and (\ref{order}) (and also for the Ashketar-Wheeler-DeWitt equation \cite{hkr}): there is simply no room for such ambiguities in the exact uncertainty approach.

The minimal interpretation underlying the exact uncertainty approach is that the configuration of a {\it physical} system is an inherently imprecise notion at a fundamental level, requiring the configuration to be modelled by a statistical ensemble in general (eg, $P(x)$, 
$P[h_{ij}]$).  Moreover, the nature of this intrinsic uncertainty is such as to preclude a classical deterministic relationship between the configuration and its conjugate momentum:  one must introduce nonclassical fluctuations into this relationship (eg, $p=\nabla S+f$).  However, the {\it degree} of indeterminism {\it is} precisely determined, at the ensemble level, being directly specified by the configuration uncertainty (eg, $\overline{f.f}=\alpha(x,P,\nabla P)$).   Note that this interpretation is consistent with, and a significant sharpening of, the statistical interpretation of quantum mechanics \cite{ballentine}.  Some comparisons with other interpretations have been made elsewhere \cite{hallreg, hkr}.

As might be expected, the exact uncertainty principle may be exemplified via exact uncertainty {\it relations}, that quantify the precise connection between the momentum fluctuations and configuration statistics at the ensemble level.  These relations have been studied in detail for quantum ensembles of particles 
\cite{hallreg, eur} and bosonic fields \cite{bosoniceur}.  For the case of a one-dimensional quantum particle they take the form 
\begin{equation} \label{eur}	
\delta x \,\Delta f = \hbar/2 
\end{equation}
for {\it all} pure states, where $\delta x$ denotes a measure of position uncertainty from classical statistics called the Fisher length \cite{eur}.  This relation is far stronger than (and implies) the usual Heisenberg uncertainty relation $\Delta x\Delta p\geq\hbar/2$.  It can equivalently be expressed as an operator relation in the standard quantum formalism, where $\Delta f$ is replaced by the uncertainty in the optimal estimate of the momentum from the measurement of position on a known state \cite{eur, bosoniceur}.

The existence of an exact uncertainty principle (and exact uncertainty relations), holding for the configuration and momentum uncertainties of all states at all times, is reminiscent of various `exact resonance' conditions in semiclassical quantum mechanics.  For example, one can obtain the `correct' energy levels of a hydrogen atom by allowing only classical orbits which contain an exact number of deBroglie wavelengths, and the `correct' energy levels of a harmonic oscillator by allowing only classical orbits with an action given by an integer multiple of $\hbar$.  In contrast, the exact uncertainty principle is fully dynamical in nature, holding for both stationary and nonstationary states, and, moreover, is `correct' for {\it all} quantum systems.

Note finally that the formalism of ensembles on configuration space, based on the assumption of an action principle for the probability density, is in general quite useful for discussing classical and quantum ensembles on equal terms - whether or not one is actually interested in `deriving' one type of ensemble from the other, and whether or not one uses the exact uncertainty approach to do so.  For example, this formalism makes it quite clear that the classical $\hbar\rightarrow 0$ limit of the 
Schr\"{o}dinger equation is a classical ensemble, rather than a classical particle.  The formalism also allows for the treatment of classical and quantum constraints on an equal footing \cite{hkr,super}.
\\
{\bf Acknowledgments}\\
I thank the organisers for the invitation to present this paper at the Fourth Australasian Conference on General Relativity and Gravitation, and am grateful to Marcel Reginatto for many valuable and enjoyable discussions on various aspects of exact uncertainty.

\end{document}